\documentclass{aa}  

\usepackage{graphicx}
\usepackage{txfonts}
\usepackage{amsmath}
\usepackage{xcolor,colortbl}
\usepackage{soul}
\usepackage{xspace}
\usepackage{enumerate}
\usepackage{placeins}
\usepackage{makecell}
\usepackage[caption=false]{subfig}
\usepackage{pdflscape}
\usepackage{rotfloat}
\usepackage{adjustbox}
\usepackage{booktabs}
\usepackage{makecell}
\usepackage{multirow}
\usepackage{natbib}
\bibpunct{(}{)}{;}{a}{}{,}
\usepackage[pdftex,bookmarks=true,colorlinks=true,linkcolor=blue,citecolor=blue, urlcolor=blue,breaklinks]{hyperref}      
\usepackage{scalerel}
\usepackage{pgfplots}
\usepackage{tikz}
\usepackage{float}
\usepackage{caption}

\usetikzlibrary{svg.path}
\usetikzlibrary{shapes.geometric}

\definecolor{orcidlogocol}{HTML}{A6CE39}
\tikzset{
  orcidlogo/.pic={
    \fill[orcidlogocol] svg{M256,128c0,70.7-57.3,128-128,128C57.3,256,0,198.7,0,128C0,57.3,57.3,0,128,0C198.7,0,256,57.3,256,128z};
    \fill[white] svg{M86.3,186.2H70.9V79.1h15.4v48.4V186.2z}
                 svg{M108.9,79.1h41.6c39.6,0,57,28.3,57,53.6c0,27.5-21.5,53.6-56.8,53.6h-41.8V79.1z M124.3,172.4h24.5c34.9,0,42.9-26.5,42.9-39.7c0-21.5-13.7-39.7-43.7-39.7h-23.7V172.4z}
                 svg{M88.7,56.8c0,5.5-4.5,10.1-10.1,10.1c-5.6,0-10.1-4.6-10.1-10.1c0-5.6,4.5-10.1,10.1-10.1C84.2,46.7,88.7,51.3,88.7,56.8z};
  }
}

\newcommand\orcidicon[1]{\href{https://orcid.org/#1}{\mbox{\scalerel*{
\begin{tikzpicture}[yscale=-1,transform shape]
\pic{orcidlogo};
\end{tikzpicture}
}{|}}}}
\bibpunct{(}{)}{;}{a}{}{,}            

\pgfplotsset{compat=1.17} 

\definecolor{cadmiumred}{rgb}{0.89, 0.0, 0.13}

\makeatletter
\renewcommand*\aa@pageof{, page \thepage{} of \pageref*{LastPage}}
\makeatother

\begin{document}

   \title{Classification of blazars based on data-driven approaches}

   \author{Simone~Vaccaro \inst{1,\orcidicon{0009-0009-6526-4828}}\thanks{Corresponding author: \texttt{prime98b@gmail.com}}
          \and
          Maria~Isabel~Carnerero \inst{2, \orcidicon{0000-0001-5843-5515}} \and
          Claudia~M.~Raiteri \inst{2, \orcidicon{0000-0003-1784-2784}} \and
          Massimo~Brescia \inst{3,1,\orcidicon{0000-0001-9506-5680}} \and
          Ylenia~Maruccia \inst{3,\orcidicon{0000-0003-1975-6310}} \and     Natale~De~Bonis \inst{1,\orcidicon{0009-0001-1547-6648}} \and     
          Giuseppe~Riccio \inst{3,\orcidicon{0000-0001-7020-1172}} \and
          Stefano~Cavuoti \inst{3,4,\orcidicon{0000-0002-3787-4196}} 
          }

   \institute{Department of Physics ``E. Pancini'', University Federico II of Napoli, Via Cinthia 21, I-80126 Napoli, Italy
         \and
             INAF, Osservatorio Astrofisico di Torino, via Osservatorio 20, I-10025 Pino Torinese, Italy
        \and
            INAF - Astronomical Observatory of Capodimonte, Via Moiariello 16, I-80131 Napoli, Italy
        \and
              INFN section of Naples, via Cinthia 6, I-80126, Napoli, Italy
             }

   \date{Received 9 February 2026; Accepted 21 May 2026}

  \abstract 
   {Active galactic nuclei (AGNs), including blazars, exhibit distinctive variability in their optical light curves, making them ideal for classification studies. This work uses data from the latest GAIA and Pan-STARRS data releases to analyze these patterns.}
   {The goal of this work is to classify AGNs into two categories: "blazars" and "non-blazars'' using only optical light curves.
   This strategy differs from most existing works, as it relies exclusively on optical variability without employing any other multiwavelength information.}
   {We processed optical light curves from GAIA and Pan-STARRS using the FATS library to extract standard time-series features. We computed additional features with custom algorithms based on literature methods. A Light Gradient-Boosting Machine (LightGBM) model was trained to classify AGNs into blazars and non-blazars based on these features. We then used this knowledge base to carry out a self-learning experiment with AGN candidates of an unknown nature.}
   {The LightGBM model achieved an accuracy of $86\%$, with precision, recall, and F1 score above 80-85\% for classifying blazars and non-blazar AGNs using optical data. The application of a BoostBoruta algorithm for feature selection reduced the feature space from 70 to 13. while maintaining comparable performance. A self-training classifier yielded similar results $85\%$, confirming the robustness of the model and the reliability of pseudo-labeling for unknown objects.}
   {}

   \keywords{Galaxies: active -- Methods: data analysis     -- Methods: statistical 
               }

   \maketitle

\section{Introduction}
Blazars are a subclass of active galactic nuclei (AGNs) hosting relativistic jets oriented close to our line of sight. This geometry causes their emission to be strongly Doppler-beamed, making them appear exceptionally bright and variable across the entire electromagnetic spectrum, from the radio band to $\gamma$ rays \citep[see eg., ][for a review]{raiteri2025}. 
The term blazar combines two main observational classes: BL Lac objects (BL Lacs) and flat-spectrum radio quasars (FSRQs). According to the original definition, BL Lacs have featureless spectra or spectra presenting emission lines with equivalent width less than 5 \AA\ in the rest frame \citep{stickel1991,stocke1991}. In contrast, FSRQs usually show broad emission lines typical of quasars.

The broadband spectral energy distributions (SEDs) of blazars is dominated by nonthermal jet emission. Both subclasses display compact radio morphologies and flat radio spectra, while their SEDs exhibit a characteristic double-humped structure.
The low-energy bump, extending from the radio to the X-ray band depending on the blazar type, is widely attributed to synchrotron radiation by ultrarelativistic electrons in the jet. The high-energy component, covering the X-ray and $\gamma$-ray energy range, could originate either from inverse-Compton scattering of the synchrotron photons (synchrotron self-Compton, SSC) or from up-scattering of external photons (external Compton, EC) originating from the accretion disk, the broad-line region (BLR), or the dusty torus. 
Overall, SSC models are commonly adopted for BL Lacs, while the EC scenario is better at describing FSRQs, whose SEDs are typically characterized by strong Compton dominance \citep{ghisellini1998}. 
Actually, the high-energy emission can also be obtained though hadronic processes, which could also be responsible for some of the high-energy neutrinos observed by neutrino telescopes, such as IceCube \citep{ice2018, Ding2023_BlNeutrini}.

The synchrotron peak frequency is used to further divide blazars into low-, intermediate-, and high-synchrotron peaked sources (LSP, ISP, and HSP, respectively; \citealt{RelJetsHovatta}):

\begin{itemize}
\item LSP : $\log \nu_\text{peak} < 14$,
\item ISP : $14 < \log \nu_\text{peak} < 15$,
\item HSP : $\log \nu_\text{peak} > 15$.
\end{itemize}

In FSRQs, the synchrotron peak typically lies in the infrared regime, making them almost exclusively LSPs, whereas in BL Lacs, it spans a wider range, from infrared up to hard X-rays. 

The selection of blazars in time-domain optical sky surveys can, in principle, exploit the extreme variability properties of these objects, which are attributed to the fact that Doppler beaming decreases the variability timescales and amplifies the variability amplitude, making blazars the most variable AGNs.
However, optical observations alone are often insufficient for a robust classification, since the jet emission can be contaminated by additional components, such as the steady host galaxy light, and the moderately variable thermal emission from the accretion disk and BLR. These components are generally present in AGNs and when they are providing an important contribution to the source emission, they ultimately challenge its classification as a blazar on a purely optical variability basis.
In contrast, these limitations are absent in other frequency bands, particularly in the $\gamma$-ray band, where the observed emission originates entirely from the jet. 
However, $\gamma$-ray light curves are available only for the brightest sources and a $\gamma$-ray detection alone is not sufficient to firmly establish the blazar nature of a source. 
Such objects are typically labeled as blazar candidates until an optical counterpart is identified and the classification verified through spectroscopy.
A rigorous classification of blazar candidates remains challenging due to the difficulty of obtaining extensive spectroscopic follow-up, the uncertainty in associating $\gamma$-ray sources with their optical counterparts, and our currently limited understanding of their intrinsic properties. 

An alternative approach consists of building and analyzing broadband SEDs, although this requires extensive multiwavelength campaigns, which are time-consuming and observationally expensive. For these reasons, the development of alternative classification strategies for blazars within the broader AGN population is of growing importance.

In recent years, machine learning has emerged as a powerful tool for blazar classification, driven by the increasing availability of data and the development of reliable algorithms. So far, most of this progress has been achieved using $\gamma$-ray observations, particularly those from the Fermi-LAT surveys. Several studies have employed machine learning to classify blazar candidates into BL Lacs and FSRQs based on high-energy data \citep[e.g][]{BL_Class_Agarwal23, BL_Class_Bhatta24}.

The goal of this work is to classify AGNs into two categories: ``blazars" and ``non-blazars'' (hereafter simply referred to as AGNs), solely using optical light curves. Our approach leverages variability information combined with boosting methods as the basis for this classification. 
A useful tool for this purpose is the Feature Analysis for Time Series (FATS)\footnote{FATS  \href{http://isadoranun.github.io/tsfeat/FeaturesDocumentation.html}{documentation}} Python library,
which enables the extraction of a wide set of statistical features from light curves \citep{FATS_doc}. These features range from simple descriptors, such as amplitude or mean magnitude, to more advanced indicators, including autocorrelation metrics and flux percentiles. Such feature-based representations provide a compact way of capturing the variability patterns of different types of variable sources \citep[e.g.][]{Richards_2011}, including AGNs and blazars, and they are well suited for machine learning applications.

We stress that the primary advantage of pursuing classification through optical time series data lies in their abundance and accessibility through large-scale, time-domain surveys such as the Zwicky Transient Facility\footnote{Zwicky Transient Facility \href{https://www.ztf.caltech.edu/}{website}} (ZTF, \citealt{2019PASP..ztf}) and the Panoramic Survey Telescope \& Rapid Response System\footnote{Pan-STARRS1 data archive \href{https://outerspace.stsci.edu/spaces/PANSTARRS/overview}{website}} (Pan-STARRS, \citealt{panstarrs1surveys}). In addition, in the next future, the Legacy Survey of Space and Time at the Vera C. Rubin Observatory\footnote{Rubin-LSST \href{https://rubinobservatory.org/}{website}} (Rubin-LSST, \citealt{2019ApJ...873..LSST}) will also provide extensive light curve datasets for blazar and non-blazar AGN classifications.

This paper is structured as follows.
Section~\ref{sec:dataset} briefly describes the dataset.
In Section~\ref{sec:method} we outline the methods adopted in this work, while Section~\ref{sec:experiments} describes our experiments. Finally, we discuss the findings and limitations in Section~\ref{sec:discussion}, and in Section~\ref{sec:conclusions} we conclude the paper and outline directions for future research.

\section{Dataset}
\label{sec:dataset}
All data used in this work were sourced from the latest data releases of the Gaia and Pan-STARRS surveys \citep{GaiaDR3_2023, panstarrs1surveys} in the form of light curves. The Gaia DR3 light curves are in the broad G band and cover the time interval from 2014 to 2017, with each curve consisting of at least 20 data points. While Gaia also observed in the blue (BP) and red (RP) bands, these are too noisy for variability studies of faint objects such as the vast majority of AGNs considered here and, thus, they were not used. The Pan-STARRS multiband data, available from 2010 to 2014, have relatively sparse sampling; therefore, we used composite light curves that were obtained by combining measurements across all filters, using average color indices \citep{2026A&A...708A.220C}. We considered the blazars in the Roma-BZCAT catalog\footnote{Roma-BZCAT \href{https://www.ssdc.asi.it/bzcat/}{catalog}}, which contains 3561 sources selected in different ways \citep{massaro2009,massaro2015}, most of which have been spectroscopically confirmed.

We looked for the blazars present in Roma-BZCAT in both Gaia and Pan-STARRS surveys, finding about 2500 objects in each of them. Since we want to test the ability of our machine learning methods to distinguish blazars from other variable AGNs, we randomly selected roughly the same number of AGNs from each survey to ensure balanced representation. For this purpose, we considered the AGNs included in the first Gaia catalog of variable AGNs \citep{GaiaDR3_2023}. In particular, we obtained two datasets:
\begin{itemize}
    \item Gaia. $2,500$ AGNs, $1,021$ BL-Lac objects, and $1,556$ FSRQs;
    \item Pan-STARRS. $2,519$ AGNs, $899$ BL-Lac objects, and $1,554$ FSRQs.
\end{itemize}
In the following, the objects in the above datasets will be referred to as ``KNOWN`` sources. We also considered the $21,733$ AGN candidates from the same Gaia catalog that lack a counterpart in other AGN catalogs and are thus taken as ``UNKNOWN`` AGNs.

\begin{figure}[!htbp]
    \centering
    \includegraphics[width=\linewidth]{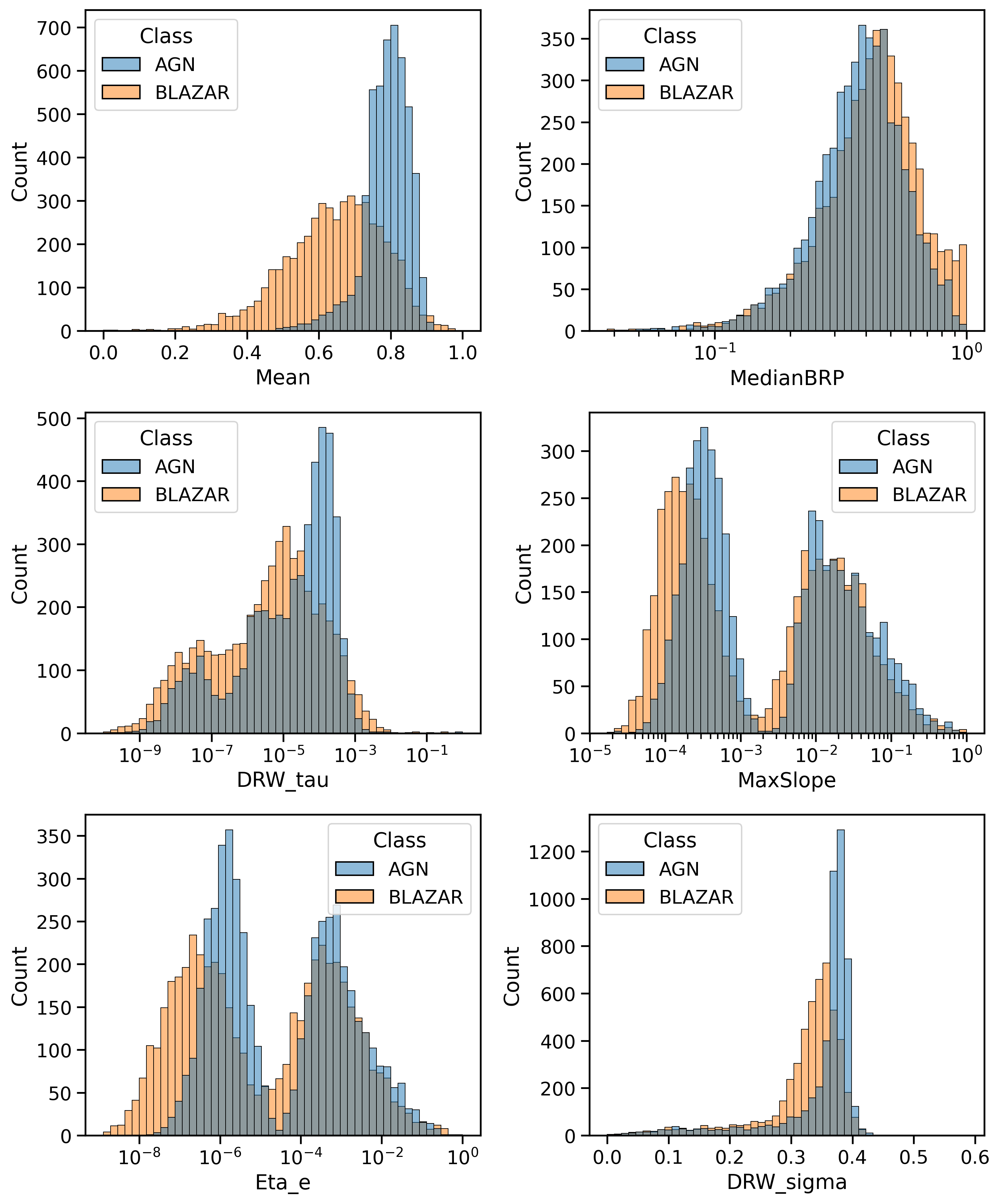}
    \caption{Distributions of selected features from the known dataset are shown on a logarithmic scale, except for \emph{Mean} and \emph{DRW\_sigma}, which are displayed on a linear scale to improve visualization. Some features, such as \emph{Mean}, show good class separation, whereas others, such as \emph{MedianBRP}, exhibit no separation.}
    \label{fig:dataset_short}
\end{figure}

The features of our time-series were extracted from the light curves using the FATS library \citep{FATS_doc}. Additionally, we computed supplementary features using custom algorithms based on literature methods. The full list of the adopted features together with their description is given in Appendix \ref{app:features} in Table~\ref{tab:variability_features}.
A full representation of our dataset can be seen in Figure ~\ref{fig:dataset}, showing the distribution of the feature values for the three source classes ``AGN," ``BLAZAR," and ``UNKNOWN." All the features were normalized using the Min-Max scaler from Python's scikit-learn library to ensure consistent scales for model training. A selected list of features are described in Table \ref{tab:variability_features_short} and some of them from the known dataset can be seen in Figure \ref{fig:dataset_short}.

\section{Method}
\label{sec:method}
For our experiment, we employed supervised and semi-supervised machine learning techniques suitable for handling labeled and unlabeled data, including Light Gradient-Boosting Machine (LightGBM) and SelfTrainingClassifier (STC). Given the large number of features, we also investigated the impact of selecting a subset of them using the BoostBoruta algorithm to verify classification performances with a simplified parameter space.

\subsection{LightGBM}

\begin{table*}[!htbp]
    \centering
    \caption{List of certain features used in this study}
    \begin{tabular}{|p{4.8cm}|p{9.5cm}|}
        \hline
        \textbf{Feature} & \textbf{Description} \\
        \hline
        $\gamma_{\mathrm{SF}}$ & Logarithmic gradient of the mean change in magnitude \\
        GP\_DRW\_$\tau$ & Relaxation time (i.e., the time necessary for the time series to become uncorrelated), from a damped random walk (DRW) model \\
        GP\_DRW\_$\sigma$ & Amplitude variability in magnitudes of the time series at short timescales ($t \ll \tau$), from a DRW model \\
        ExcessVar & Measure of the intrinsic variability amplitude \\
        CAR\_mean & CAR is a continuous time auto regressive model with three parameters: mean, sigma and tau \\
        CAR\_sigma & CAR variance\\
        CAR\_tau & CAR characteristic time\\
        $\eta^e$ & Ratio of the mean of the squares of successive mag differences to the variance of the light curve \\
        FluxPercentileRatioMid20,35, 50, 65, 80 & Percentile ratios. Where $F_{i,j}$ is the difference between $i\%$ and  $j\%$ magnitude values. FPRMid20 is $F_{40,60}/F_{5,95}$\\
        Gskew & Median-based measure of the skew \\
        MaxSlope & Maximum absolute magnitude slope between two consecutive observations \\
        Mean & Mean magnitude \\
        MedianBRP & Fraction of photometric points within amplitude/10 of the median magnitude \\
        StetsonK & Robust kurtosis measure \\
        \hline
    \end{tabular}
    \tablefoot{Most of them are calculated with the FATS python library, while the others were custom-implemented in Python based on methods from the literature.}
    \label{tab:variability_features_short}
\end{table*}

\begin{figure}[!htbp]
    \centering
    \includegraphics[width=0.49\textwidth]{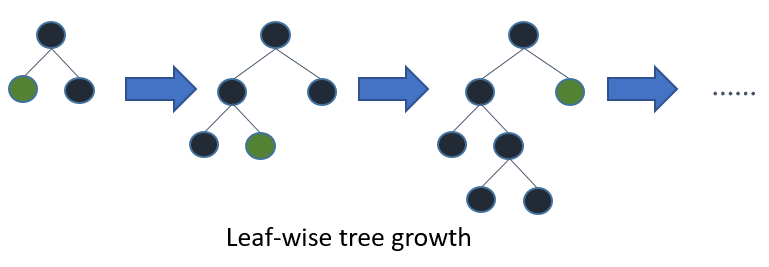}
    \caption{Gradient-boosting decision tree with leaf-wise growth, as in LightGBM, which prioritizes leaves with maximum delta loss for lower overall loss \citep{LightGBM_Paper2017}.}
    \label{figML:GBDT decision trees}
\end{figure}

We employed a machine learning approach based on the LightGBM algorithm \citep{LightGBM_Paper2017} to explore the performance of a binary AGN-blazar classification based on optical variability only, while overcoming the limitations of traditional methods that require multiband data and spectroscopic follow-up. LightGBM is a gradient-boosting decision tree model with a leaf-wise growth strategy.
The leaf-wise strategy grows trees by choosing the leaf that yields the greatest reduction in the loss function at each step. This approach, illustrated in Figure~\ref{figML:GBDT decision trees}, often achieves lower losses, compared to level-wise methods for a fixed number of leaves.

LightGBM optimizes computational efficiency through:
\begin{itemize}
    \item Gradient-based one-side sampling (GOSS) retains instances with large gradients, which contribute significantly to information gain, while randomly sampling those with small gradients and improving efficiency without sacrificing accuracy;
    \item Exclusive feature bundling (EFB) gathers mutually exclusive features in sparse feature spaces, reducing the number of effective features while preserving split point accuracy.
\end{itemize}
The model was trained on a balanced dataset of blazars and AGNs, with performance evaluated using five-fold cross-validation to ensure robust generalization (see Section~\ref{sec:experiments} for further details).

The model performance was evaluated using precision (purity), recall (completeness), F1 score, and accuracy:
\begin{itemize}
    \item Precision is the fraction of predicted blazars that are correctly classified;
    \item Recall is the fraction of known blazars that have been correctly classified;
    \item F1 score is the harmonic mean of precision and recall, balancing false positives and negatives;
    \item Accuracy is the overall fraction of correct classifications.
\end{itemize}

\subsection{Self-learning classifier}
\begin{figure}[!htbp]
    \centering
    \includegraphics[width=1\linewidth]{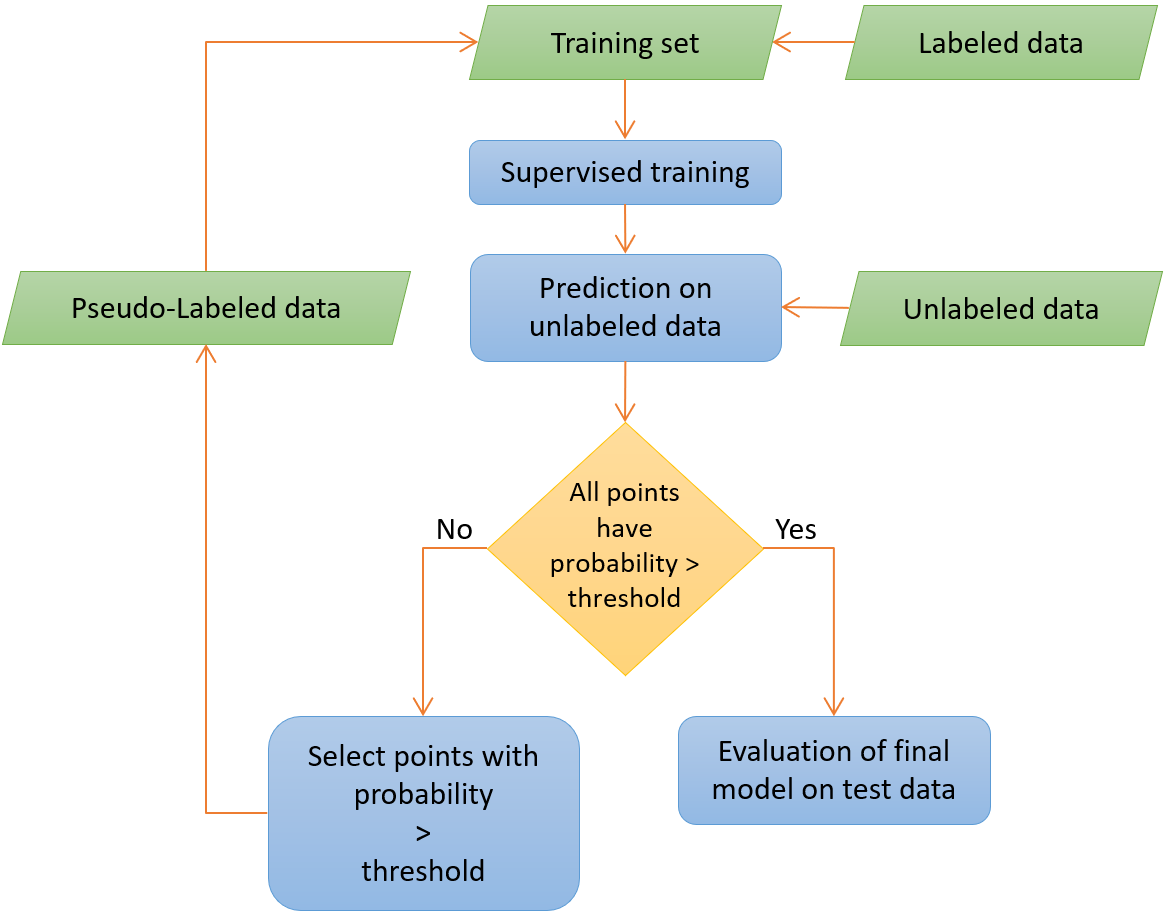}
    \caption{Workflow of our SLC. The pseudo-labeling cycle continues until all unknown objects are pseudo-labeled and then we perform a final test.}
    \label{fig:SelfTrain_scheme}
\end{figure}

To extend the classification to unlabeled data, we conducted a semi-supervised learning experiment by adding 21,733 unknown objects to the dataset. Self-training models \citep{Self_train_Class_doc} iteratively assign pseudo-labels to unlabeled objects based on a classifier’s predictions, incorporating high-confidence predictions into the training set to refine the model at each iteration. 
In detail, the self-training process begins with an initial supervised classifier (a surrogate model) trained only on the known portion of the dataset. At each iteration, this classifier generates probabilistic predictions for the unlabeled objects. Pseudo-labels are then assigned only to those unlabeled instances where the model's confidence exceeds a predetermined threshold.
The newly pseudo-labeled objects are appended to the training set, and the classifier is retrained on this expanded dataset. This iterative process continues until convergence criteria are met, such as when all unlabeled objects have been assigned pseudo-labels with sufficient confidence, no further high-confidence predictions are available, or a maximum number of iterations is reached (to prevent overfitting to noisy labels).
Figure~\ref{fig:SelfTrain_scheme} illustrates the workflow of the self-training process. The results of both the binary classification and semi-supervised experiments are reported in Section~\ref{sec:discussion}.

\subsection{Feature selection with BoostBoruta and SHAP}
The feature selection was performed using the BoostBoruta package\footnote{Software available  \href{https://github.com/cerlymarco/shap-hypetune}{here}}. The Boruta algorithm \citep{BorutaPaper} identifies relevant features by creating randomized copies (the features' values are shuffled), known as shadow features, and iteratively comparing their importance to that of the original features using an internal classification, traditionally based on random forests. Features consistently outperforming their shadow counterparts are deemed significant. BoostBoruta extends this approach by replacing the standard random forest model with gradient boosting, enhancing robustness and efficiency.
For the binary classification case (blazars vs. AGNs), BoostBoruta incorporates SHapley Additive exPlanations (SHAP) values \citep{SHAP_Paper} to enhance the selection of relevant features. The SHAP values quantify each feature’s contribution to predictions by considering all possible subsets of features and measuring how the prediction changes when a given feature is added.
This results in a set of additive attributions, the SHAP values, that are consistent (features that always increase the prediction receive positive contributions and vice versa) and locally accurate (the sum of the contributions equals the model output).

\section{Experiments}
\label{sec:experiments}

The first crucial step for machine learning algorithms is the division of the known sample of AGNs and blazars into training and validation subsets. To evaluate the model performance, we employed a five-fold cross-validation strategy instead of a single train-test split. In this approach, the dataset is randomly divided into 5 equal folds. For each iteration, the model is trained on four folds ($80\%$ of the data) and validated on the remaining fold ($20\%$ of the data). This process is repeated five times, ensuring that every data point is used for both training and validation exactly once. Performance metrics, defined in section~\ref{sec:method}, were averaged across all folds to provide an unbiased estimate of the model’s generalization to unseen data.
This five-fold cross-validation was chosen to maximize the use of the available data, particularly given the relatively small size of the original known dataset, while providing a more robust assessment of the model's reliability and stability compared to a single 80:20 split. It helps mitigate variability in performance estimates and ensures that the model can be aptly generalized across different subsets of the data.

\subsection{Classification with LightGBM on a known dataset}
\label{subsec:lgbm_exp}

Experiments on the known dataset were conducted with the LightGBM classifier\footnote{LightGBM Package  \href{https://lightgbm.readthedocs.io/en/latest/pythonapi/lightgbm.LGBMClassifier.html}{documentation}}, both with and without feature selection to evaluate its impact on model performance. A hyperparameter tuning was performed once for both experiments (with and without feature selection) prior to five-fold cross-validation using scikit-learn’s GridSearchCV function\footnote{GridSearchCV package \href{https://scikit-learn.org/stable/modules/generated/sklearn.model_selection.GridSearchCV.html}{documentation}}. This step streamlines the typically complex manual process of hyperparameter optimization.
We optimized \texttt{learning\_rate}, \texttt{n\_estimators}, and \texttt{colsample\_bytree}, while keeping other hyperparameters at their default values.
The \texttt{learning\_rate} controls the step size of weight updates during the boosting process. A smaller value enhances robustness by making gradual adjustments to predictions, improving generalization at the cost of requiring more iterations, whereas a larger value accelerates convergence but risks overshooting optimal solutions. The \texttt{n\_estimators} parameter determines the number of boosted trees constructed, with higher values enabling the model to capture complex patterns but potentially leading to overfitting if not carefully tuned. The \texttt{colsample\_bytree} parameter governs the fraction of features randomly sampled for each tree, introducing randomness to reduce overfitting and enhance generalization. These hyperparameters were chosen for optimization because they directly influence the balance between model complexity, training efficiency, and predictive performance, making them critical for achieving robust and well-calibrated models.
The ROC-AUC\footnote{(Receiver Operating Characteristic Area Under the Curve): quantifies the model’s ability to distinguish between classes across varying decision thresholds} score was used as optimization metric due to its reliability for binary classification tasks.
Further technical details and hyperparameter settings of the model are reported in Appendix ~\ref{app:setup}.

The model was trained on the labeled training set of each fold, and performance metrics were computed on the corresponding test set to establish a baseline for comparison with the semi-supervised approach.

The feature importance, based on the number of splits in LightGBM, was computed for each fold and averaged to identify the most influential features and they are included with mean importance values. To further investigate how the model exploits the available information and to assess the role of individual features in the classification process, we complemented the standard feature importance analysis with a SHAP-based interpretation for the LightGBM experiment.
In particular, we considered both local and global explanations of the model predictions. At the local level, we used a SHAP waterfall plot to illustrate how individual features combine to drive a specific prediction. At the global level, we employed SHAP beeswarm plots to visualize the overall impact and distribution of feature contributions across the dataset, accounting for the k-fold cross-validation procedure.
The results of this SHAP analysis are presented in Section \ref{sec:Feat_SHAP}.

\subsection{Self-training classification with an unknown dataset}
In the second experiment, we focused on the unknown sources in our dataset. For this purpose we tried to capitalize the knowledge base with the known dataset and try to give a classification to the unknown objects.
We employed STC\footnote{Self-training classifier package \protect\href{https://scikit-learn.org/stable/modules/generated/sklearn.semi_supervised.SelfTrainingClassifier.html\#sklearn.semi_supervised.SelfTrainingClassifier}{documentation}} based on Yarowsky’s algorithm \citealt{Self_train_Class_doc}) with the same LightGBM base estimator as in the supervised experiment.
Two configurations were tested: one using all available features and another with feature selection applied using the same features selected by BoostBoruta, consistent with the LightGBM experiment. 
For both configurations, we used the same hyperparameters of the previous experiment, mirroring the optimized parameters from the supervised case.
For each fold of the five-fold cross-validation, the training set was augmented by combining the labeled training data ($80\%$ of the labeled dataset) with the unlabeled dataset, where unlabeled samples were assigned a placeholder label of $-1$, as required by the self-training algorithm. A confidence threshold of $0.95$ was chosen for assigning pseudo-labels to unlabeled data during iterative training, balancing the accuracy of pseudo-labels with the inclusion of unlabeled samples while minimizing the propagation of errors. 
STC was trained until all pseudo-labels had been assigned to unlabeled samples based on the model's confidence. Predictions were made on the same test sets used in the LightGBM experiment, ensuring a direct comparability of performance metrics. Further technical details and hyperparameter settings of the model are reported in Appendix ~\ref{app:setup}.

Feature importance, based on the number of splits in the LightGBM based estimator, was computed for each fold and averaged to identify the most influential features for the STC, highlighting their role in the semi-supervised classification.

\section{Results and discussion}
\label{sec:discussion}

The experiments used a dataset of $10,089$ labeled objects, with STC incorporating $21,733$ unlabeled objects (the dataset is described in Section \ref{sec:dataset}). 
Tables~\ref{tab:lgbm_metrics} and Table~\ref{tab:stc_metrics} summarize the performance metrics across five-fold cross-validation, respectively, for LightGBM and STC. 
Both models were tested with and without feature selection through BoostBoruta, illustrated in Table \ref{tab:feat_sel}, to assess their impact on performance and feature importance. Figures~\ref{fig:lgbm_features} and~\ref{fig:stc_features} illustrate the top ten feature importance values for LightGBM and STC, respectively.
All the aforementioned tables and figures are reported and discussed in detail in the following subsections.

\begin{table}[!hbpt]
    \centering
    \caption{Feature selection with the BoostBoruta model}
    \begin{tabular}{|l|}
    \hline
    Feature selected\\
    \hline
    CAR\_mean\\
    CAR\_sigma\\
    CAR\_tau\\
    Eta\_e\\
    FluxPercentileRatioMid20\\
    Gskew\\
    MaxSlope\\
    Mean\\
    PercentAmplitude\\
    StetsonK\\
    gamma\\
    DRW\_sigma\\
    DRW\_tau\\
    \hline
    \end{tabular}
    \tablefoot{The BoostBoruta model is used for both the LightGBM and STC experiments. The method reduces the parameter space from 70 features to 13.}
    \label{tab:feat_sel}
\end{table}

\subsection{LightGBM classification} 
\label{sec:lgbm_class_diss}

The LightGBM model, trained with the full set of approximately $70$ features, achieved an accuracy of $86\%$, with precision, recall, and F1 score exceeding 80-85\% for both classes (Table~\ref{tab:lgbm_metrics}), despite the strong overlap between AGN and blazar classes in the statistical features derived from optical data. Promisingly, these results were achieved without data from other bands (radio, infrared, X-ray, and $\gamma$-ray), which are commonly used for blazar classification, highlighting the potential of our approach in addressing this complex challenge.

\begin{figure}[!htbp]
\centering
\includegraphics[width=1\linewidth]{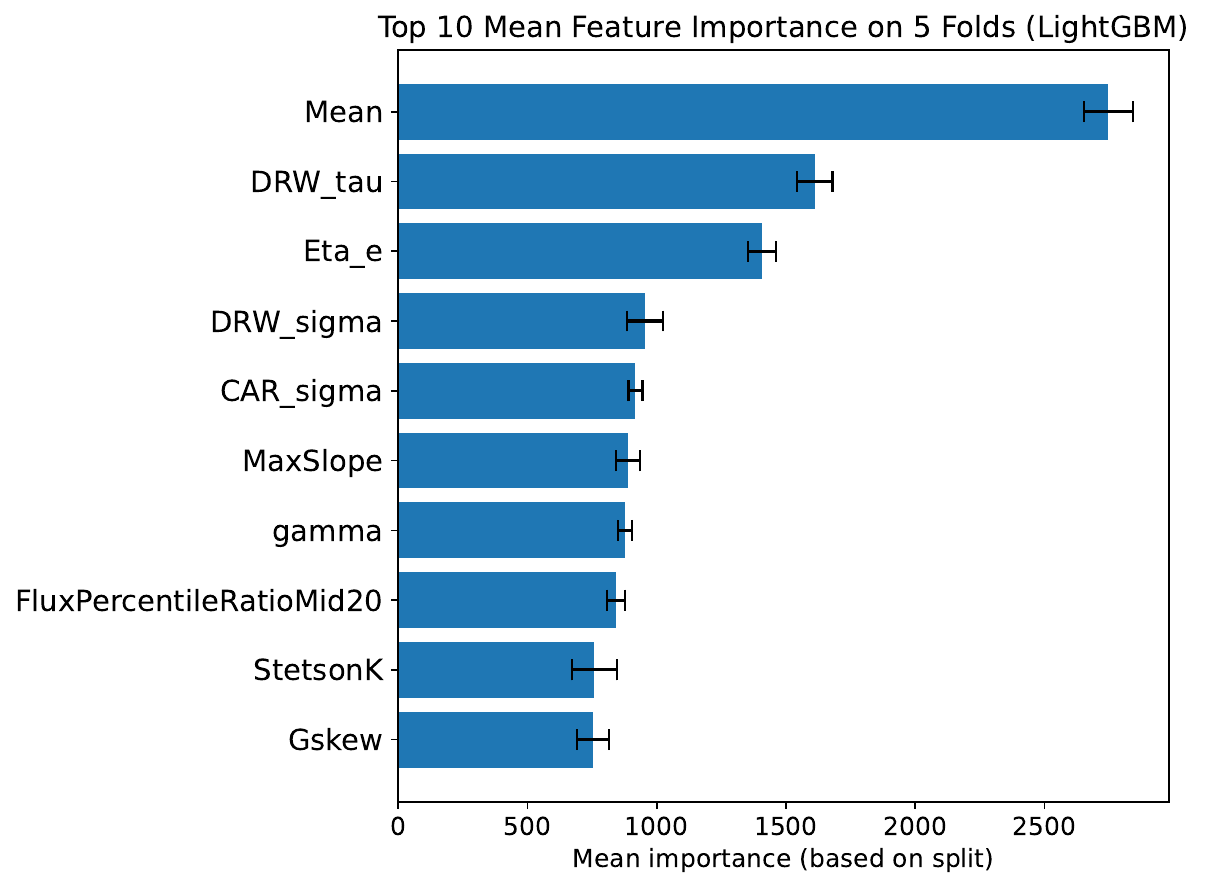}
\caption{Top ten feature importance values (mean $\pm$ standard deviation) for the LightGBM classifier across a five-fold cross-validation.}
\label{fig:lgbm_features}
\end{figure}

The feature selection using the BoostBoruta algorithm reduced the feature space to $13$ key features, while maintaining performance comparable to the full feature set. This result highlights the algorithm's ability to identify an optimal subspace, achieving an effective balance between model simplicity and predictive power. Figure~\ref{fig:lgbm_features} illustrates the top ten features ranked by mean importance ($\pm$ standard deviation) across five-fold cross-validation for the LightGBM model. Key features, such as \emph{Mean}, \emph{Eta\_e}, and \emph{DRW\_tau}, exhibited high importance, reflecting their critical role in distinguishing AGNs from blazars. The consistency of influential features between the full model and the BoostBoruta-selected subset, coupled with a significant reduction in parameter space, confirms the robustness of the feature selection process. This is further evidenced by the clear separation in importance values shown in Figure~\ref{fig:lgbm_features}, emphasizing the dominance of a small subset of features in addressing this classification.

\begin{table}[!htbp]
\centering
\caption{LightGBM Results}
\begin{tabular}{|l|c|c|}
\hline
\textbf{Metric} & \textbf{LightGBM (All)}& \textbf{LightGBM (FS)}\\
\hline
Accuracy           & 0.857 $\pm$ 0.012 & 0.856 $\pm$ 0.012  \\
Precision (AGN)    & 0.818 $\pm$ 0.018 & 0.817 $\pm$ 0.018  \\
Recall (AGN)       & 0.913 $\pm$ 0.012 & 0.912 $\pm$ 0.010  \\
F1 (AGN)           & 0.863 $\pm$ 0.012 & 0.862 $\pm$ 0.012  \\
Precision (BLAZAR) & 0.904 $\pm$ 0.012 & 0.903 $\pm$ 0.011  \\
Recall (BLAZAR)    & 0.803 $\pm$ 0.016 & 0.802 $\pm$ 0.015  \\
F1 (BLAZAR)        & 0.850 $\pm$ 0.011 & 0.849 $\pm$ 0.011  \\
\hline
\end{tabular}
\tablefoot{Performance metrics (mean $\pm$ standard deviation) with and without feature selection, across a five-fold cross-validation.}
\label{tab:lgbm_metrics}
\end{table}

\subsection{Self-learning classification} 
To further evaluate the model's robustness, we conducted an experiment using an STC and incorporating an unknown class. The STC achieved an accuracy of $85\%$, with the precision, recall, and F1 score exceeding $80-85\%$ for both classes, as shown in Table~\ref{tab:stc_metrics}, closely matching the performance of the baseline LightGBM model. This consistency is notable, as the STC iteratively assigns pseudo-labels to unknown objects during training, demonstrating the reliability of these pseudo-labels.

\begin{figure}[!htbp]
\centering
\includegraphics[width=1\linewidth]{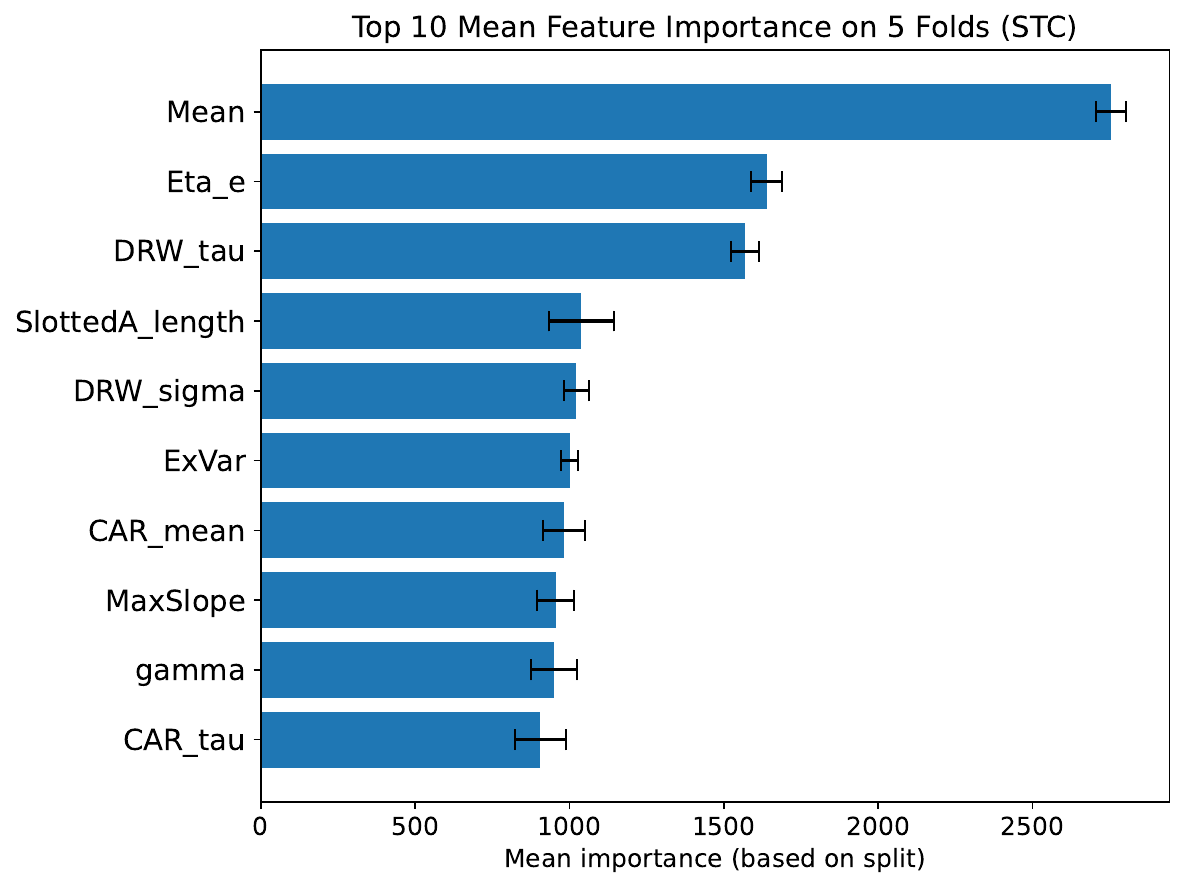}
\caption{Top ten feature importance values (mean $\pm$ standard deviation) for the STC classifier across the five-fold cross-validation.}
\label{fig:stc_features}
\end{figure}

The experiment with the BoostBoruta-selected feature set (13 features) yielded a comparable performance, reinforcing the effectiveness of the feature selection process even with the unknown objects. 
Figure~\ref{fig:stc_features} presents the top ten features ranked by mean importance ($\pm$ standard deviation) across the five-fold cross-validation for the STC model, with key features such as \emph{Mean}, \emph{Eta\_e}, and \emph{DRW\_tau} remaining prominent, consistent with the LightGBM results. This agreement in feature importance, combined with the successful handling of the expanded dataset including unknown objects, underscores the stability and robustness of the model. The results confirm that the selected features are highly effective for both classification and pseudo-labeling tasks, as evidenced by the clear separation in importance values in Figure~\ref{fig:stc_features}.

\begin{table}[!htbp]
\centering
\caption{Results for STC}
\begin{tabular}{|l|c|c|}
\hline
\textbf{Metric} & \textbf{STC (All)} & \textbf{STC (FS)}  \\
\hline
Accuracy           & 0.854 $\pm$ 0.013 & 0.853 $\pm$ 0.010  \\
Precision (AGN)    & 0.814 $\pm$ 0.020 & 0.814 $\pm$ 0.020  \\
Recall (AGN)       & 0.914 $\pm$ 0.012 & 0.911 $\pm$ 0.004  \\
F1 (AGN)           & 0.861 $\pm$ 0.014 & 0.856 $\pm$ 0.011  \\
Precision (BLAZAR) & 0.905 $\pm$ 0.012 & 0.902 $\pm$ 0.005  \\
Recall (BLAZAR)    & 0.796 $\pm$ 0.017 & 0.797 $\pm$ 0.018  \\
F1 (BLAZAR)        & 0.847 $\pm$ 0.012 & 0.846 $\pm$ 0.010  \\
\hline
\end{tabular}
\tablefoot{Performance metrics (mean $\pm$ standard deviation) with and without feature selection, across the five-fold cross-validation.}
\label{tab:stc_metrics}
\end{table}

\subsection{Most important features and analysis}
\label{sec:Feat_SHAP}
In all the experiments performed in this study, the three features—\emph{Mean}, \emph{Eta\_e}, and \emph{DRW\_tau}—are by far the most important for classification, as they are consistently selected by the BoostBoruta method and exhibit the highest rankings in feature importance plots (Figures \ref{fig:lgbm_features} and \ref{fig:stc_features}). This prominence likely arises from both technical and physical factors, which are inherently interconnected. From a technical perspective, these features show more separation between class distributions compared to all others (Figure \ref{fig:dataset_short}, Figure \ref{fig:dataset} in Appendix \ref{app:features} for all distributions), enabling models to rely on them more effectively for classification. This separation, however, reflects underlying physical differences. 
The \emph{Mean}, the top-ranked feature, corresponds to the average magnitude of the source; AGNs dominated by thermal disk emission are systematically less luminous than most blazars, making the dominance of this feature expected. The other two features, \emph{Eta\_e} and \emph{DRW\_tau}, show comparable importance to each other in the rankings but consistently outperform all remaining features except the \emph{Mean}.
The \emph{Eta\_e} parameter, which measures the ratio of the mean squared successive magnitude differences to the light-curve variance, is a direct probe of variability timescale; its importance arises from the distinct variability behavior of the blazar class, which exhibits more intense and rapid fluctuations (on timescales of even hours), whereas disk-dominated AGNs display smoother magnitude evolution. Similarly, \emph{DRW\_tau} (i.e., the characteristic timescale for the light curve to become uncorrelated under a damped random walk model) captures this fast-versus-slow dichotomy, with shorter values expected for blazars and longer ones for AGNs. 
Despite their physical relevance, the distributions of \emph{Eta\_e} and \emph{DRW\_tau} are seen to largely overlap, reflecting a nuanced distinction between the classes, rather than a sharp one.

\begin{figure}
    \centering
    \includegraphics[width=\linewidth]{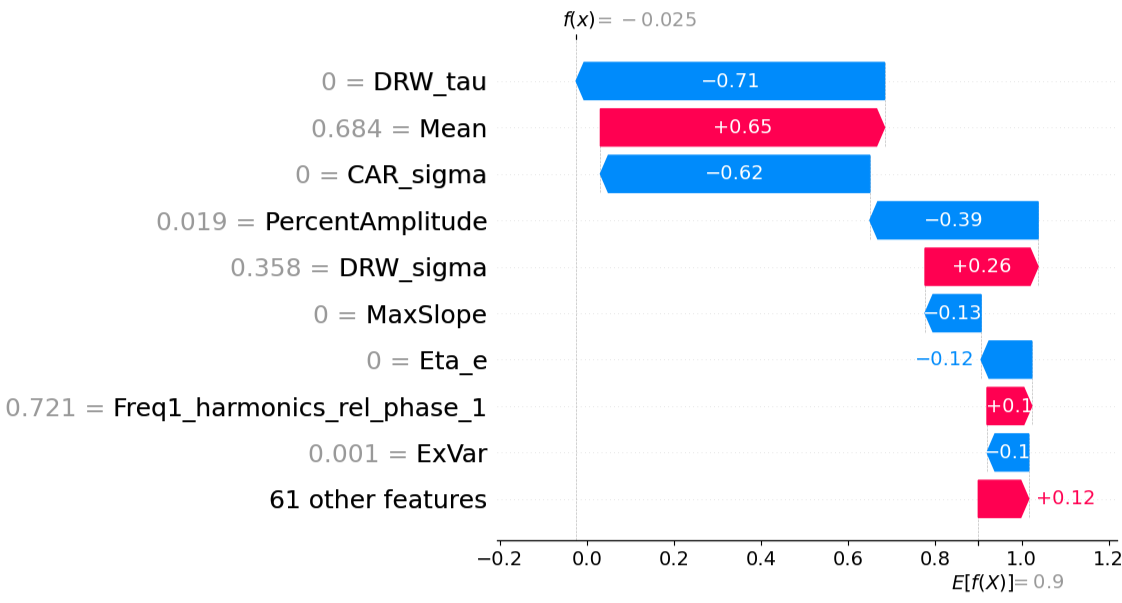}
    \caption{Waterfall plot for a false negative classification. The values in bars show how the various features contribute to classification. In our case, $f(x) < 0$ implies that the object is classified as an AGN.}
    \label{fig:Waterfall}
\end{figure}

\begin{figure}
    \centering
    \includegraphics[width=\linewidth]{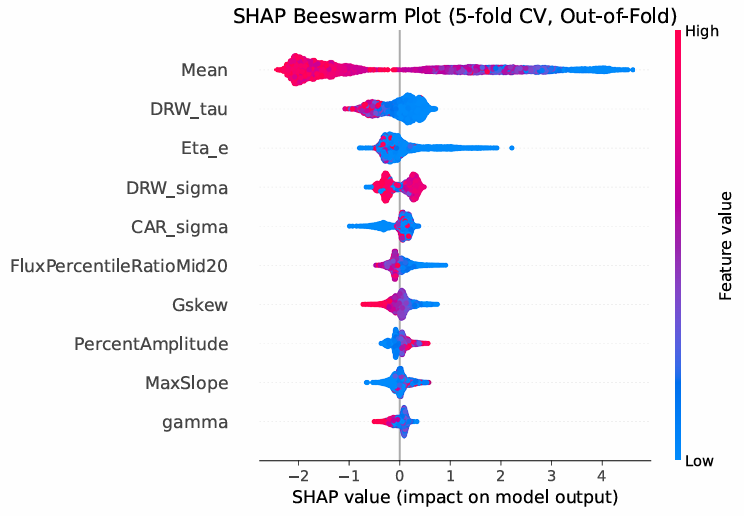}
    \caption{Beeswarm plot of data for our experiment with LightGBM. For visualization purposes, only the ten most important features are shown.}
    \label{fig:beeswarm}
\end{figure}

The analysis of features contributing to our LightGBM classification can be seen in Figure \ref{fig:Waterfall} and \ref{fig:beeswarm}, displaying a waterfall and a beeswarm plot.
For the waterfall plot, we selected a representative false negative example, as it highlights the contributions of features in borderline or ambiguous cases (Figure \ref{fig:Waterfall}). On the left side of the plot, the values of the features for the selected object are reported, while the horizontal bars represent their contributions to the model output. Starting from the expected value, each feature shifts the prediction towards higher or lower values, and the sum of these contributions yields the model output shown on top. Thus, we have $f(x)=E+\sum_i C_i$, where $C_i$ represents the feature contributions. Minor discrepancies may arise when visually summing the displayed values, as the contributions reported in the plot are rounded for visualization purposes.

In the beeswarm plot shown in Figure \ref{fig:beeswarm}, each point represents an individual object; its position along the horizontal axis indicates the SHAP value (i.e., the contribution to the model output), while its color encodes the corresponding feature value for that object. The vertical spread reflects the density of points with similar contributions. In Figure \ref{fig:beeswarm}, the feature \emph{Mean} clearly emerges as one of the most influential, as indicated by the wide range of SHAP values it spans. Higher values of \emph{Mean} are generally associated with negative SHAP values (favoring the AGN class), while lower values tend to push the prediction towards the Blazar class. However, the plot also shows that \emph{Mean} alone is not sufficient for a consistent classification: objects with similar \emph{Mean} values can still exhibit a broad range of SHAP contributions, including cases where high \emph{Mean} values are associated with positive SHAP values.

\section{Conclusions}
\label{sec:conclusions}

This study demonstrates the efficacy of machine learning in distinguishing blazars from
non-blazar AGNs using optical time series data. This result is not taken for granted since, in the optical band, the jet's very variable emission (one of the distinctive features of blazars) can be damped by the steady light of the stars of the host galaxy or by the smoothly variable emission contributions from the accretion disc and BLR; if this emission ends up prevailing over the nonthermal jet emission, it can make blazars resemble normal AGNs. Nonetheless, the methods we applied led to a robust discrimination among them. 

The supervised LightGBM model achieved an accuracy of $86\%$, while the semi-supervised STC reached $85\%$, with precision, recall, and F1 scores exceeding $80-85\%$ for both classes (Tables~\ref{tab:lgbm_metrics}, \ref{tab:stc_metrics}). 
The BoostBoruta algorithm reduced the feature space from approximately 70 to 13, identifying key discriminative features such as \emph{Mean}, \emph{Eta\_e}, and \emph{DRW\_tau}, while maintaining comparable performance. 
The STC’s reliable pseudo-labeling of $21,733$ unlabeled objects further confirms the model’s robustness, enabling effective classifications, despite the challenges of optical data.
These results are particularly interesting because they are based entirely on optical time series, without the need of data from other bands.
The primary advantage of optical data lies in their accessibility and volume: large-scale sky surveys such as ZTF, Pan-STARRS, or the upcoming Rubin-LSST, offer millions of light curves with dense temporal sampling, facilitating population-level studies and the inclusion of fainter sources, capturing a broader AGN population. 
Our approach, leveraging BoostBoruta and STC, capitalizes on this data abundance to achieve reliable classification, demonstrating the capacity of machine learning to overcome optical limitations.

Future works could enhance accuracy by integrating multiband data, such as colors and spectral features, to mitigate feature overlap. With the upcoming Rubin-LSST data release expected to reveal tens of thousands of new blazars \citep{Raiteri_2022LSST}, our methodology lays a foundation for efficient and robust classification in large-scale optical surveys.

\begin{acknowledgements}
      Part of this work was supported by Italian Research Center on High Performance Computing Big Data and Quantum Computing (ICSC), project funded by European Union - NextGenerationEU - and National Recovery and Resilience Plan (NRRP) - Mission 4 Component 2 within the activities of Spoke 3 (Astrophysics and Cosmos Observations).

      We acknowledge the use of the ADHOC (Astrophysical Data HPC Operating Center) resources, within the project "Strengthening the Italian Leadership in ELT and SKA (STILES)", proposal nr. IR0000034, admitted and eligible for funding from the funds referred to in the D.D. prot. no. 245 of August 10, 2022 and D.D. 326 of August 30, 2022, funded under the program "Next Generation EU" of the European Union, “Piano Nazionale di Ripresa e Resilienza” (PNRR) of the Italian Ministry of University and Research (MUR), “Fund for the creation of an integrated system of research and innovation infrastructures”, Action 3.1.1 "Creation of new IR or strengthening of existing IR involved in the Horizon Europe Scientific Excellence objectives and the establishment of networks.

      C.M.R. and M.I.C acknowledge financial support from the INAF Fundamental Research Funding Call 2023 (PI: Raiteri).
\end{acknowledgements}

\bibliographystyle{aa}
\bibliography{biblio}

\begin{appendix}

\onecolumn
\section{FATS features distributions and description}
\label{app:features}

\noindent
\begin{minipage}{\textwidth}
  \centering  
    \includegraphics[width=\linewidth]{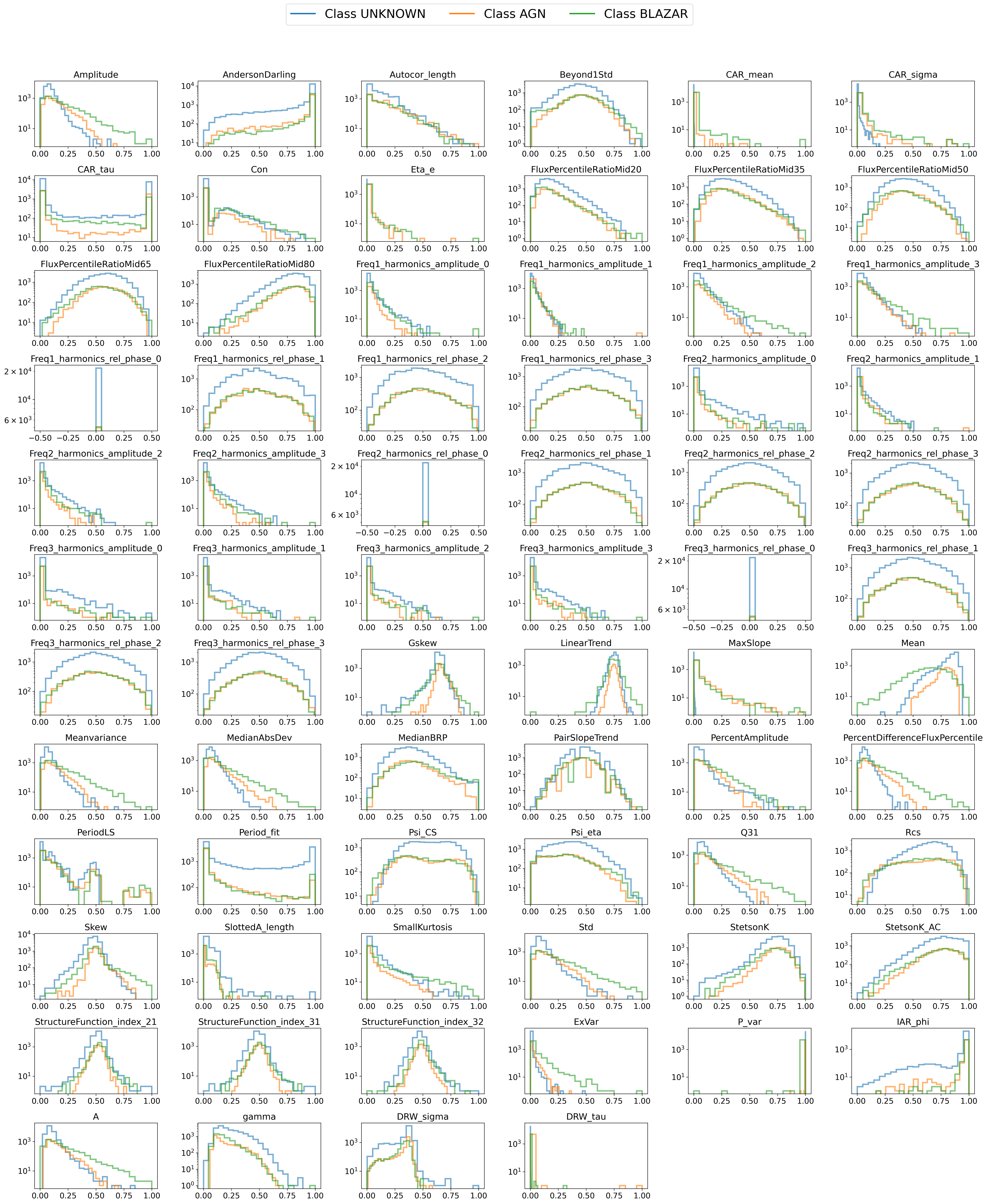}
    \captionof{figure}{Distributions of each feature of our dataset in logarithmic scale. All features were normalized to ensure consistent scales for model training.}
    \label{fig:dataset}
\end{minipage}

\begin{table*}[!htbp]
    \centering
    \caption{List of all features used in this study}
    \resizebox{.95\linewidth}{!}{
    \begin{tabular}{|p{4.8cm}|p{9.5cm}|p{3.5cm}|}
        \hline
        \textbf{Feature} & \textbf{Description} & \textbf{Reference} \\
        \hline
        \multicolumn{3}{|c|}{\textbf{Custom-implemented Features}} \\
        \hline
        $A_{\mathrm{SF}}$ & RMS magnitude difference of the Structure Function (SF), computed over a 1 yr timescale & \cite{Schmidt_2010} \\
        $\gamma_{\mathrm{SF}}$ & Logarithmic gradient of the mean change in magnitude & \cite{Schmidt_2010} \\
        GP\_DRW\_$\tau$ & Relaxation time (i.e., time necessary for the time series to become uncorrelated), from a Damped Random Walk (DRW) model & \cite{Graham_2017} \\
        GP\_DRW\_$\sigma$ & Amplitude variability in magnitudes of the time series at short timescales ($t \ll \tau$), from a DRW model & \cite{Graham_2017} \\
        ExcessVar & Measure of the intrinsic variability amplitude & \cite{Allevato_2013} \\
        $P_{\mathrm{var}}$ & Probability that the source is intrinsically variable & \cite{1996ApJ...473..763M} \\
        IAR$_\phi$ & Level of autocorrelation using a discrete-time representation of a DRW model & \cite{Eyheramendy_2018} \\
        \hline
        \multicolumn{3}{|c|}{\textbf{FATS Features}} \\
        \hline
        Amplitude & Half of the difference between the median of the maximum 5\% and of the minimum 5\% magnitudes & \cite{Richards_2011} \\
        AndersonDarling & Test of whether a sample of data comes from a population with a specific distribution & \cite{FATS_doc} \\
        Autocor\_length & Lag value where the autocorrelation function becomes smaller than $\eta^e$ & \cite{Kim_2011} \\
        Beyond1Std & Percentage of points with photometric mag that lie beyond $1\sigma$ from the mean & \cite{Richards_2011} \\
        CAR\_mean & CAR is a continuous time auto regressive model with three parameters: mean, sigma and tau & \cite{Pichara_2012} \\
        CAR\_sigma & The CAR variance & \cite{Pichara_2012} \\
        CAR\_tau & The CAR characteristic time & \cite{Pichara_2012} \\
        Con & It is defined by the count of the number of three consecutive data points that are above or below $2\sigma$ from the mean magnitude and normalize this number by $N-2$. & \cite{Kim_2011} \\
        $\eta^e$ & Ratio of the mean of the squares of successive mag differences to the variance of the light curve & \cite{Kim_2014} \\
        FluxPercentileRatioMid20,35, 50, 65, 80 & Percentile ratios. Where $F_{i,j}$ is the difference between $i\%$ and  $j\%$ magnitude values. FPRMid20 is $F_{40,60}/F_{5,95}$ & \cite{Richards_2011}\\
        Freq\{i\}\_harmonics\_amplitude\_\{j\} & Amplitudes of the harmonics extracted from light curves using Lomb–Scargle periodogram & \cite{Richards_2011}\\
        Freq\{i\}\_harmonics\_rel\_phase\_\{j\} & Phases of the harmonics extracted from light curves using Lomb–Scargle periodogram & \cite{Richards_2011}\\
        Gskew & Median-based measure of the skew & ---- \\
        LinearTrend & Slope of a linear fit to the light curve & \cite{Richards_2011} \\
        MaxSlope & Maximum absolute magnitude slope between two consecutive observations & \cite{Richards_2011} \\
        Mean & Mean magnitude & \cite{FATS_doc}\\
        Meanvariance & Ratio of the standard deviation to the mean magnitude & \cite{FATS_doc} \\
        MedianAbsDev & Median discrepancy of the data from the median & \cite{Richards_2011} \\
        MedianBRP & Fraction of photometric points within amplitude/10 of the median magnitude & \cite{Richards_2011} \\
        PairSlopeTrend & Fraction of increasing first differences minus decreasing ones over the last 30 time-sorted mag measures & \cite{Richards_2011} \\
        PercentAmplitude & Highest percentage difference between either max or min mag and median mag & \cite{Richards_2011} \\
        PercentDifferenceFluxPercentile & Ratio of the difference between the $95\%$ and the $5\%$ over the median magnitude & \cite{Richards_2011}\\
        Period\_LS & Period derived from the Lomb-Scargle method & \cite{Kim_2011}\\
        Period\_fit & False-alarm probability of the largest periodogram value obtained with LS & \cite{Kim_2011} \\
        $\Psi_{\mathrm{CS}}$ & Range of a cumulative sum applied to the phase-folded light curve & \cite{Kim_2014} \\
        $\Psi_{\eta}$ & $\eta^e$ index calculated from the folded light curve & \cite{Kim_2014} \\
        Q31 & Difference between the third and the first quartile of the light curve & \cite{Kim_2014} \\
        $R_{\mathrm{CS}}$ & Range of a cumulative sum & \cite{Kim_2011} \\
        Skew & Skewness measure & \cite{Kim_2011} \\
        SlottedA\_lenght & As Autocor\_length, but time lags are defined as intervals or slots instead of single values. & \cite{Protopapas_2015}\\
        SmallKurtosis & Kurtosis of the light curve. & \cite{Richards_2011} \\
        Std & Standard deviation of the light curve & \cite{FATS_doc} \\
        StetsonK & Robust kurtosis measure & \cite{Kim_2011} \\
        \hline
    \end{tabular}}
    \tablefoot{Most of them are calculated with the FATS python library, while the others are custom-implemented in Python based on methods from the literature.}
    \label{tab:variability_features}
\end{table*}

\twocolumn
\section{Model configuration and hyperparameters}
\label{app:setup}
All experiments were performed on a desktop PC with an 8-core, 16-thread CPU. Each model fitting and classification run lasted only a few seconds.

\subsection{LightGBM classifier}
We employed the LightGBM classifier for our experiments, both with and without feature selection to evaluate its impact on model performance. 
Specifically, we optimized \texttt{learning\_rate} $\in[0.005, 0.05]$, \texttt{n\_estimators} $\in[300, 1200]$, and \texttt{colsample\_bytree} $\in[0.3, 1]$. 

We set \texttt{objective} = 'binary', \texttt{random\_state} $=42$, \texttt{verbose} $= -1$, and \texttt{n\_jobs} $=8$, while keeping other hyperparameters at their default values. The \texttt{random\_state} was fixed to ensure reproducibility across k-fold cross-validation runs and \texttt{n\_jobs} was set equal to the number of physical CPU cores to maximize efficiency, as recommended in the LightGBM classifier documentation.

The ROC-AUC score was used as the optimization metric due to its robustness in binary classification tasks.

The optimal hyperparameters for the experiment with all features were:
\begin{itemize}
    \item \texttt{learning\_rate} = $0.0075$,
    \item \texttt{n\_estimators} = $1,200$,
    \item \texttt{colsample\_bytree} = $0.5$;
\end{itemize}
while those for the experiment with feature selection:
\begin{itemize}
    \item \texttt{learning\_rate} = $0.0075$,
    \item \texttt{n\_estimators} = $1,000$,
    \item \texttt{colsample\_bytree} = $0.6$.
\end{itemize}

The model was trained on the labeled training set of each fold, and performance metrics were computed on the corresponding test set to establish a baseline for comparison with the semi-supervised approach.

\subsection{SelfTrainingClassifier}
STC was configured using the same LightGBM model as the base estimator.

We set \texttt{threshold} $=0.95$, \texttt{criterion} = 'threshold', and \texttt{max\_iter} = 'None'.
The \texttt{threshold} value of $0.95$ was chosen to balance the accuracy of pseudo-labels with the inclusion of unlabeled samples, while minimizing error propagation. The \texttt{max\_iter} parameter was set to 'None' to train the STC until all pseudo-labels were assigned to the unlabeled samples.

Two configurations were tested: one using all available features and another with feature selection applied using the same subset of features selected by BoostBoruta, consistent with the LightGBM experiments.
For both configurations, the LightGBM base estimator was initialized using the same set of optimized hyperparameters defined above.

\end{appendix}
\end{document}